\documentclass[review]{elsarticle}
\usepackage{lineno,hyperref}
\usepackage[english]{babel}
\usepackage[numbers]{natbib}
\usepackage{xcolor}
\usepackage{latexsym,amsmath,amssymb,amsbsy,graphicx,geometry}
\modulolinenumbers[5]

\begin{document}
	
	\begin{frontmatter}
	
\title{Machine learning-based prediction of elastic properties of amorphous metal alloys}

\author[kfu,ufrc]{Bulat N. Galimzyanov\corref{cor1}}
\cortext[cor1]{Corresponding author}
\ead{bulatgnmail@gmail.com}

\author[kfu]{Maria A. Doronina} 

\author[kfu,ufrc]{Anatolii V. Mokshin}
\ead{anatolii.mokshin@mail.ru} 

\address[kfu]{Kazan Federal University, 420008 Kazan, Russia} 
\address[ufrc]{Udmurt Federal Research Center of the Ural Branch of RAS, 426067 Izhevsk, Russia}

\begin{abstract}
The Young's modulus $E$ is the key mechanical property that determines the resistance of solids to tension/compression. In the present work, the correlation of the quantity $E$ with such characteristics as the total molar mass $M$ of alloy components, the number of components $n$ forming an alloy, the yield stress $\sigma_{y}$ and the glass transition temperature $T_{g}$ has been studied in detail based on a large set of empirical data for the Young's modulus of different amorphous metal alloys. It has been established that the values of the Young's modulus of metal alloys under normal conditions correlate with such a mechanical characteristic as the yield stress as well as with the glass transition temperature. As found, the specificity of the ``chemical formula'' of alloy, which is determined by molar mass $M$ and number of components $n$, does not affect on elasticity of the material. The machine learning algorithm identified both the quantities $M$ and $n$ as insignificant factors in determining $E$. A simple non-linear regression model is obtained that relates the Young's modulus with $T_{g}$ and $\sigma_{y}$, and this model correctly reproduces the experimental data for metal alloys of different types. This obtained regression model generalizes the previously presented empirical relation $E\simeq49.8\sigma_{y}$ for amorphous metal alloys.
\end{abstract}


\end{frontmatter}

\section{Introduction}

Knowledge of the Young's modulus $E$ plays a crucial role in the development of functional materials based on amorphous metal alloys and in ensuring the mechanical stability of structures in various industries~\cite{Pelleg_book_2013,Preetha_Sreekala_2018,Cavaliere_2021}. The modulus $E$ characterizes the elasticity of a solid material and its resistance to external mechanical influences. From a microscopic point of view, elasticity is defined by interatomic (intermolecular) interaction forces as well as structure of a solid~\cite{Zhuang_Liu_2012,Dai_Yu_Dong_2022,Galimzyanov_Doronina_2022}. In this regard, the elastic properties of the material must be determined by the type of chemical elements that form this material as well as by the ratio of the concentrations of these elements, i.e., in what proportions on relation to each other the different elements are presented in the material~\cite{Wang_2006,Torres_Stafford_2010}. However, there are no simple relationships connecting these quantities to each other. The solution of this problem within the framework of the microscopic theory of elasticity is nontrivial.

In the case of amorphous solids, many of their physical (mainly, mechanical) properties are determined by the so-called glass forming ability (GFA)~\cite{Binder_Kob_2005,Wang_Angell_2006,Chahal_Ramesh_2022,LouzguineLuzgin_2022,Baggioli_Zaccone_2022}. In other words, any amorphous solid is characterized by some GFA, which determines the ability of the corresponding melt to form an amorphous phase. Although there is no generally accepted GFA criterion, practically all known GFA criteria are determined through the glass transition temperature $T_{g}$~\cite{Li_2001,Masood_Belova_2020, Galimzyanov_Mokshin_viscosity_2021}. In this regard, it seems quite reasonable to expect that the glass transition temperature $T_g$ can be related to the elastic properties of amorphous solids. Thus, in the case of amorphous solids, in particular, amorphous metal alloys, we formulate the following question: is there a correspondence between the elastic modulus and some physical and chemical characteristics of solids, such as the yield stress $\sigma_{y}$, the molar mass $M $, the number of components $n$ forming the material, and the glass transition temperature $T_{g}$?

Young's modulus can be determined from the well-known ``stress-strain'' relation, which indicates correspondence between strain and stress for a material~\cite{Beghini_Bertini_2006,Galimzyanov_Mokshin_NiTi_2021}. Here, the magnitude of $E$ is determined from the slope of the initial linear part of the ``stress-strain'' relation corresponding to the elastic region~\cite{Clickner_Ekin_2006,Arrayago_Gardner_2015}. Sometimes, it becomes necessary to estimate $E$ from known physical characteristics without mechanical tests and computer simulations. This task is especially relevant in the case of fragile materials with a complex sample preparation procedure, in the case of synthesizing new materials with required mechanical properties, and in the case of predicting strength characteristics of existing materials under various thermodynamic conditions. It is noteworthy that modern artificial intelligence methods, in particular, machine learning, can be used to solve this task~\cite{Liu_Zhao_2022,Mokshin_Mirziyarova_2020,Klimenko_Ryltsev_2022,Mokshin_Khabibullin_2022,Balyakin_Yuryev_2022}.

Machine learning methods have been repeatedly applied to predict the value of the Young's modulus of various materials including metal alloys~\cite{Khakurel_Taufque_2021}, silicate glasses~\cite{Yang_Xu_2019}, polymeric materials~\cite{Pugar_Gang_2022} and rocks~\cite{Shahani_Zheng_2022}. These studies have shown the efficiency of machine learning based on the artificial neural networks (ANN's) in estimating the Young's modulus from known physical characteristics of materials. For example, in Ref.~\cite{Khakurel_Taufque_2021}, it was shown that the average concentration of valence electrons, the difference in atomic radii and the melting temperature are the key characteristics that determine the Young's modulus of refractory alloys. Despite the successes achieved, similar studies have not yet been carried out with respect to such an extensive class of materials as amorphous metal alloys.

In the present study, the calculation of the Young's modulus of amorphous metal alloys with different compositions and mechanical properties is performed by machine learning based on the ANN's. This calculation is carried out by set of experimental data for metal alloys based on Al, Au, Ca, Co, Cu, Fe, La, Hf, Mg, Ni, Pd, Pt, Re, Sr, Ti, W, Zr and rare earth elements~\cite{Qu_Liu_2015}. As an input parameters in the ANN, we chose such physical characteristics as the total molar mass $M$ of the alloy components, the number of components $n$, the yield stress $\sigma_{y}$, and the glass transition temperature $T_g$. Information about these characteristics is available in the case of the considered amorphous metal alloys. It has been established that these characteristics are sufficient to determine the Young's modulus with high accuracy.

\section{Neural network construction and learning}

\begin{figure}[h!]
	\centering
	\includegraphics[width=0.8\linewidth]{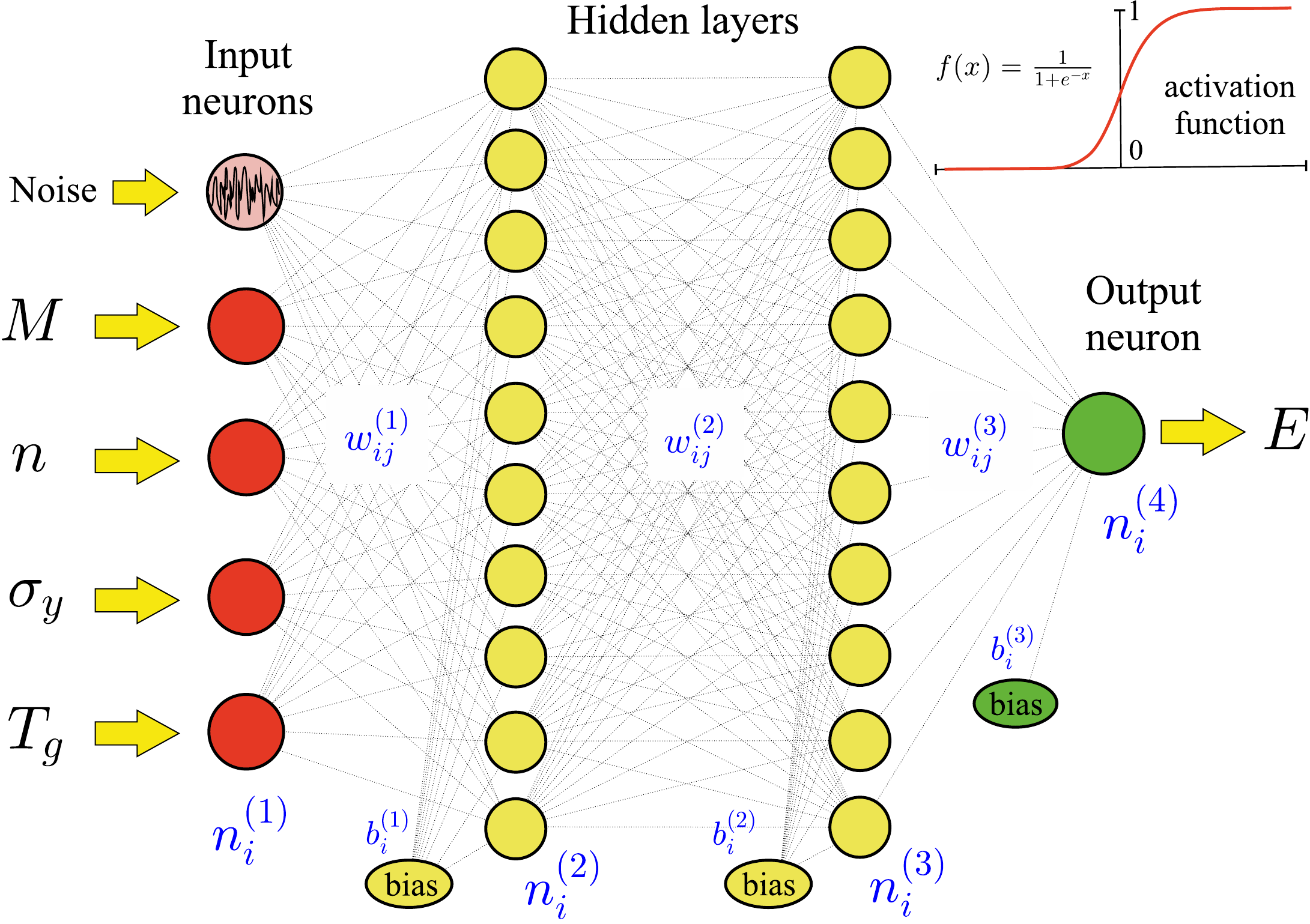}
	\caption{Scheme of the artificial neural network designed to calculate the Young's modulus $E$ of metal alloys from the known values of the molar mass $M$, the number of components $n$, the yield stress $\sigma_{y}$ and the glass transition temperature $T_{g}$, and also taking into account the noise factor (Noise).}
	\label{fig_1}
\end{figure}

In the present work, the ANN of direct propagation was built to perform the required calculations. This ANN consists of four layers (see Fig.~\ref{fig_1}). The first layer contains $5$ input neurons. The next two layers are hidden and each of them contains ten neurons. The values of the Young's modulus $E$ are determined by the one output neuron. The input neurons are supplied by values of the molar mass $M$, the number of components $n$, the yield stress $\sigma_{y}$, the glass transition temperature $T_{g}$ and the noise factor for  different metal alloys (see Tables~1--5 in Supporting Materials). All the necessary data for these alloys are taken from Ref.~\cite{Qu_Liu_2015}. The values of the input physical quantities are preliminarily calibrated so that they change in the range from $0$ to $1$. For learning and testing of the ANN, we have used metal alloys, whose the Young's modulus $E$ is known. For the remaining alloys, the values of $E$ are predicted by the ANN after it has been learned.

The main working regime of the ANN is the direct propagation of information, in which the output value of each neuron is calculated as follows~\cite{Chumachenko_Gabbouj_2022}
\begin{equation}\label{eq_ns_1}
n_{i}^{(k)}=f(w_{11}^{(k-1)}n_{1}^{(k-1)}+w_{12}^{(k-1)}n_{2}^{(k-1)}+...+w_{ij}^{(k-1)}n_{j}^{(k-1)}+b_{i}^{(k)})
\end{equation}
or
\begin{equation}\label{eq_ns_2}
n_{i}^{(k)}=f\left(W_{i}^{(k)}+b_{i}^{(k)}\right),
\end{equation}
where the input value of the $k$th layer neuron is calculated through the expression
\begin{equation}\label{eq_ns_3}
W_{i}^{(k)}=\sum_{j=1}^{N_{k-1}}w_{ij}^{(k-1)}n_{j}^{(k-1)}.
\end{equation}
Here, $n_{i}^{(k)}$ is value of the $i$th neuron in the $k$th layer ($k=1,\,2,\,3,\,4$); $w_{ij}^{(k-1)}$ is value of the weight of the $(k-1)$th layer going from the neuron with index $j$ to the neuron with index $i$ from the $k$th layer; $b_{i}^{(k)}$ is the bias weight acting on $i$th neuron from the $k$th layer; $N_{k-1}$ is the number of neurons in the $(k-1)$th layer; $f(...)$ is the sigmoid activation function
\begin{equation}\label{eq_sigma_func}
f(x)=\frac{1}{1-e^{-x}},
\end{equation}
which takes values from $0$ to $1$. The weights $w_{ij}^{(k-1)}$ and $b_{i}^{(k)}$ are assigned the fixed value $0.5$ before the first iteration of the ANN.
\begin{figure}[h!]
	\centering
	\includegraphics[width=0.7\linewidth]{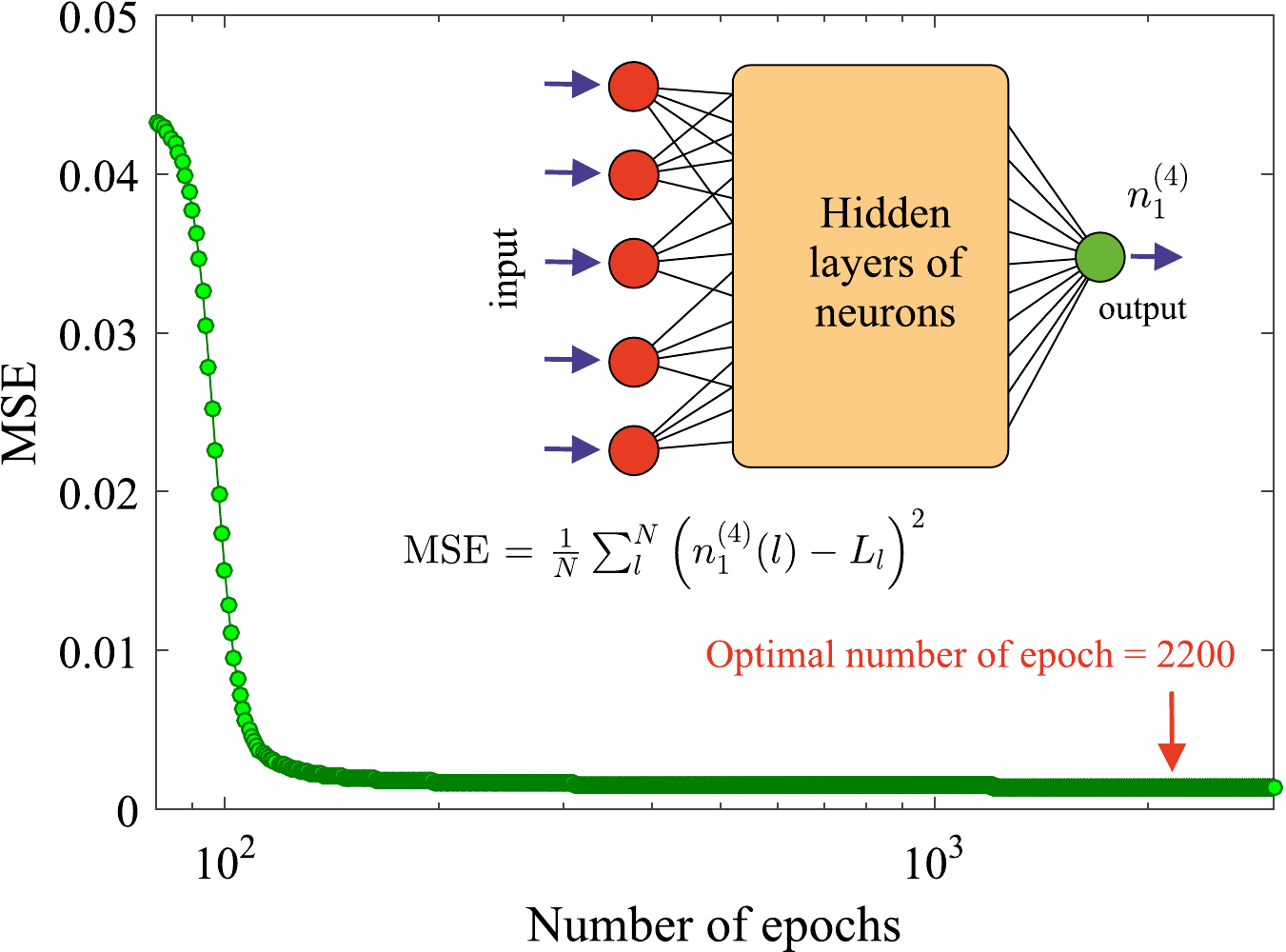}
	\caption{Neural network loss function that is calculated using Eq.~(\ref{eq_lose_function}) according to  test data set for $20$ different metal alloys.}
	\label{fig_2}
\end{figure}

For learning and minimizing the loss function of the ANN, the backpropagation and the gradient descent methods are applied~\cite{Haykin_2009}. These methods allow us to adjust the values of all weights taking into account the error between the output neuron and a required result (details are discussed in Supporting Materials in section ``1. Artificial neural network learning method''). The criterion to finish learning is the passage of a certain number of epochs. The optimal number of epochs is determined by calculation of the mean square error (MSE), which in turn is a loss function for the considered ANN:
\begin{equation}\label{eq_lose_function}
\text{MSE}=\frac{1}{N}\sum_{l=1}^{N}\left(n_{1}^{(4)}(l)-L_{l}\right)^{2}.
\end{equation}
Here, $N=20$ is the number of alloys in the test data set formed from different types of alloys. Figure~\ref{fig_2} shows that the result of Eq.~(\ref{eq_lose_function}) is the decaying curve, which start to take the minimal value MSE$\simeq0.0015$ at after $2200$ epochs. This number of epochs can be considered as optimal, since with a further increase in the number of epochs the MSE increases due to retraining of the ANN.

\section{Results}

\begin{figure}[h!]
	\centering
	\includegraphics[width=0.7\linewidth]{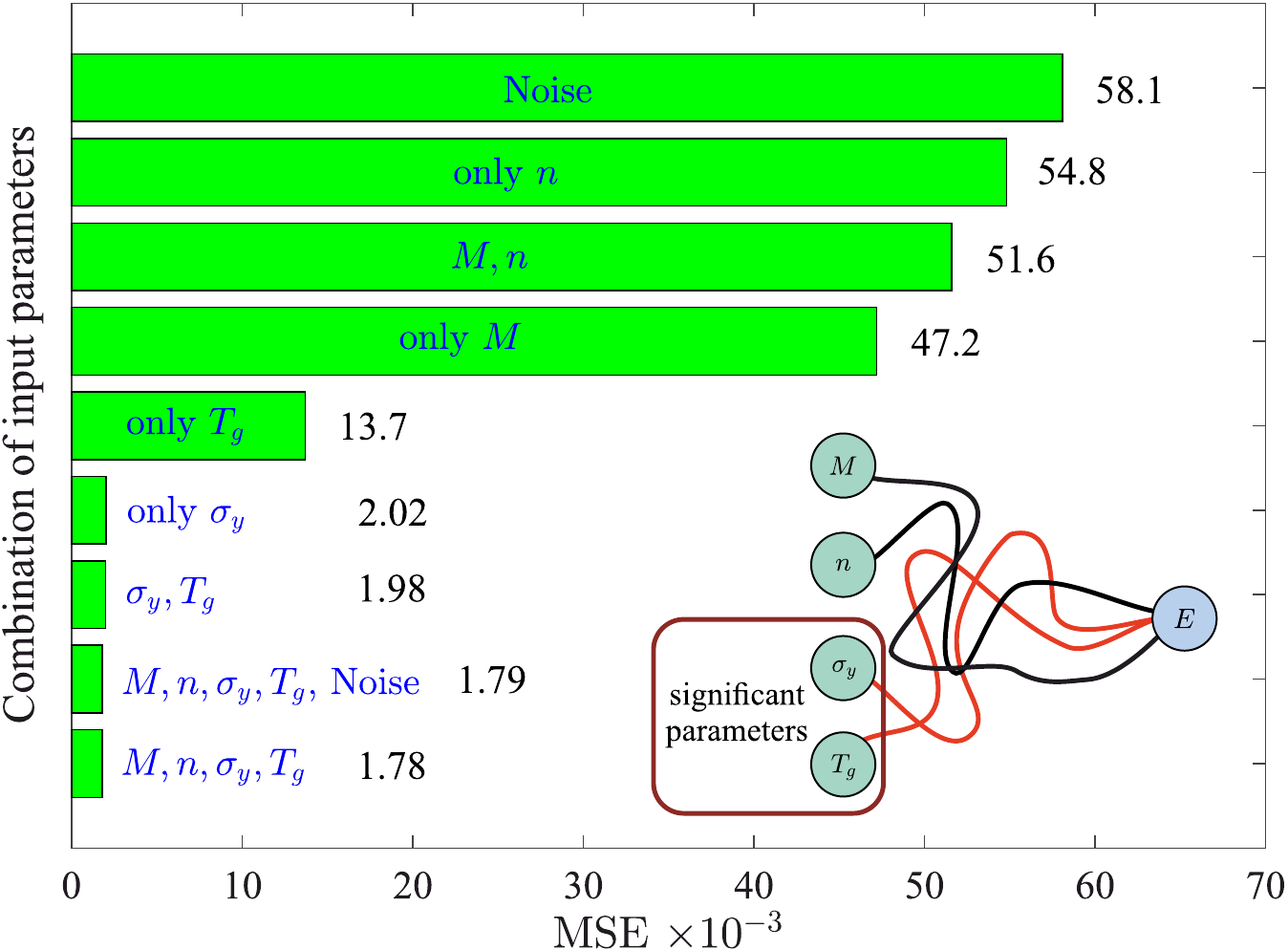}
	\caption{MSE values at various combinations of the ANN input parameters.}
	\label{fig_3}
\end{figure}

After learning the ANN, test calculations were carried out and the most significant physical parameters were identified, the values of which with the values of the Young's modulus. For this purpose, the MSE was calculated for various combinations of physical parameters at the ANN input. As can be seen from Figure~\ref{fig_3}, the smallest MSE is achieved when all four input physical parameters $M$, $n$, $\sigma_{y}$ and $T_g$ are taken into account and is equal to $\sim0.00178$. Addition of the noise factor leads to a slight increase in the error. Exclusion from consideration of the molar mass $M$ and the number of components $n$ also has no significant effect on the result and it increases the MSE up to $\simeq0.00198$. Thus, it can be seen from Figure~\ref{fig_3} that the yield stress $\sigma_{y}$ and the glass transition temperature $T_g$ are the most significant parameters. This is also confirmed by the fact that taking only $\sigma_{y}$ as an input factor leads to MSE$\simeq0.00202$, while only with $T_{g}$ the MSE is $\simeq0.0137$. Further, $M$ and $n$ are identified as insignificant factors, the MSE of which is comparable with the error of the noise factor, $\sim0.058$. This result shows that the molar mass $M$ and the number of components $n$ do not reveal a correlation with the resistance of a material to mechanical deformation.

\begin{figure}[h!]
	\centering
	\includegraphics[width=1.0\linewidth]{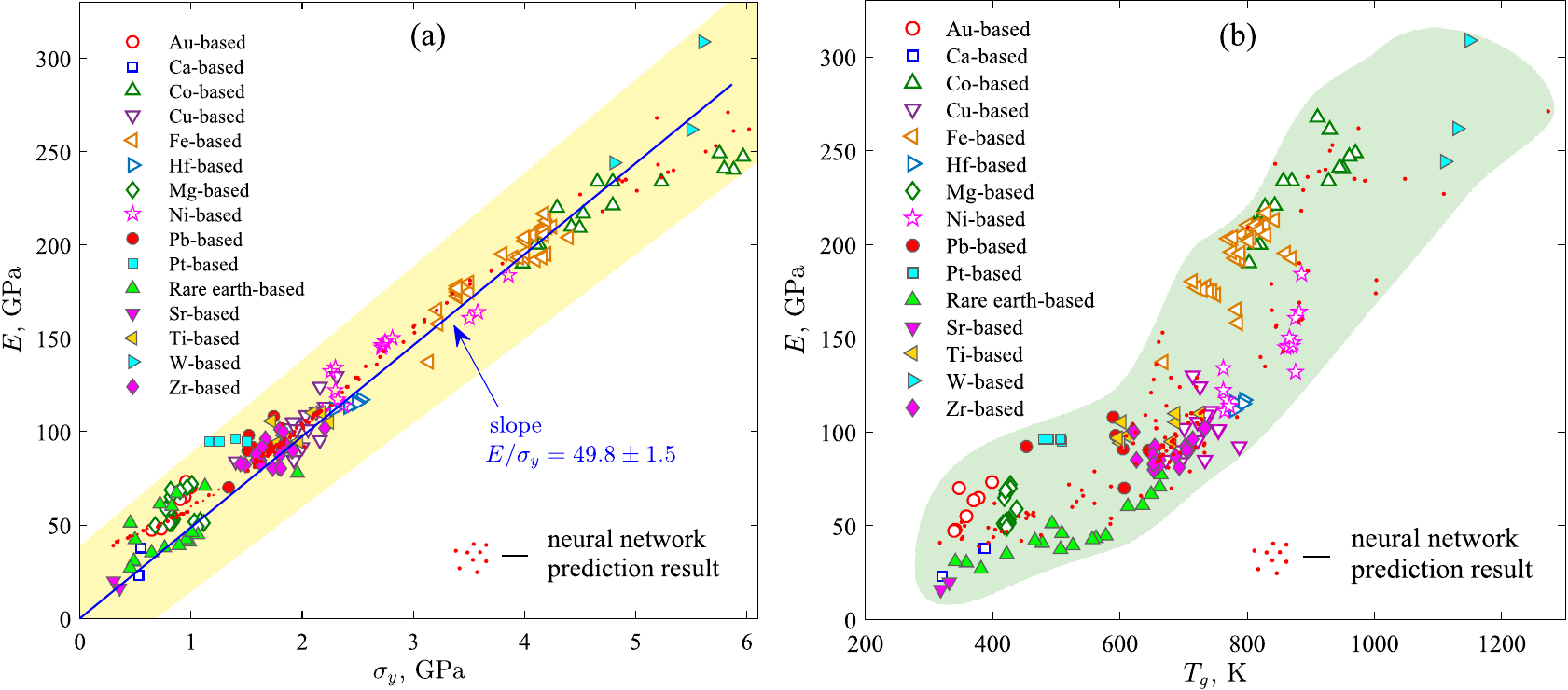}
	\caption{(a) Correlation between the Young's modulus $E$ and the yield stress $\sigma_{y}$, which obeys the linear law $E/\sigma_{y}\simeq49.8$. (b) Correspondence between the Young's modulus $E$ and the glass transition temperature $T_g$.}
	\label{fig_4}
\end{figure}

For the considered alloys, there is a clear relationship between the yield stress $\sigma_{y}$ and the Young's modulus $E$, that is confirmed by the results presented in Figure~\ref{fig_4}(a). As seen, there is a linear relationship between $E$ and $\sigma_{y}$, which can be reproduced by the function $E=49.8\sigma_{y}$. This result confirms the known relation $E/\sigma_{y}\simeq49.8$ obtained earlier in Ref.~\cite{Qu_Liu_2015} on the basis of empirical data. We have found that the ANN result fits this linear relation even for alloys whose Young's modulus was not previously known [red dots in Figure~\ref{fig_4}(a)]. This confirms the applicability of the ANN to predict the Young's modulus of various types of metal alloys based on the known values of the yield stress $\sigma_{y}$.

Figure~\ref{fig_4}(b) shows that there is no simple relationship between $E$ and $T_g$. The correspondence between these physical parameters is not reproduced by any simple law due to the fact that for a type of alloy there is a certain range of $E$ and $T_g$ values. Despite this, the prediction of the ANN fits into the resulting ranges of $E$ and $T_g$ [see Figure~\ref{fig_4}(b)], which confirms the presence of a hidden relationship between the Young's modulus and the glass transition temperature.

To quantitative characterization of the hidden relationship between $E$ and $T_g$ as well as to obtain a general equation relating the quantities $E$, $\sigma_{y}$, $T_g$, we have studied the reproducibility of data in the framework of a non-linear regression model of the following form~\cite{Mokshin_Mirziyarova_2020,Seber_Wild_1989}:
\begin{equation}\label{eq_nrm_1}
Y(X_{1},X_{2})=a_{0}+a_{1}X_{1}+a_{2}X_{2}+a_{3}X_{1}X_{2}+a_{4}X_{1}^{2}+a_{5}X_{2}^{2}+a_{6}(X_{1}X_{2})^{2}.
\end{equation}
Here, $Y\equiv E$ is the output factor, while $X_{1}\equiv\sigma_{y}$ and $X_{2}\equiv T_{g}$ are the input factors. The values of the coefficients $a_{0}$, $a_{1}$, ..., $a_{6}$ were selected by enumeration from the range [$-2000$; $2000$] with the increments  $ds=500$. For each combination of the values of these coefficients, the error value was calculated as follows
\begin{equation}\label{eq_nrm_err}
\xi=\sqrt{\frac{1}{N_{Exp}}\sum_{i=1}^{N_{Exp}}(Y_{i}-E_{i})^{2}},
\end{equation}
where $Y_{i}$ is the Young's modulus calculated from relation (\ref{eq_nrm_1}), where the  input factors $\sigma_{y}$ and $T_g$ take the experimental values. The quantity $E_{i}$ characterizes the experimental values of the Young's modulus. $N_{Exp}=173$ is the number of alloys for which the experimental values of $\sigma_{y}$, $T_{g}$, and $E$ are known (see Tables~1--5 in Supporting Materials). The values of the coefficients $a_{0}^{(best)}$, $a_{1}^{(best)}$, ..., $a_{6}^{(best )}$ giving the best fit between the experimental Young's modulus and the result of Eq.~(\ref{eq_nrm_1}) are determined at a minimal value of the error $\xi$. Further, the ranges of coefficient values are narrowed and set as [$a_{0}^{(best)}-ds$; $a_{0}^{(best)}+ds$], [$a_{1}^{(best)}-ds$; $a_{1}^{(best)}+ds$], ..., [$a_{6}^{(best)}-ds$; $a_{6}^{(best)}+ds$]. The increment $ds$ is also halved. The search for the optimal values of the coefficients is repeated anew until the step size becomes $ds<0.001$. The values of the fitting coefficients were determined as $a_{0}=25$\,GPa, $a_{1}=41.4$, $a_{2}=-0.0046$\,GPa /K, $a_{3}=0.0015$\,$K^{-1}$, $a_{4}=a_{5}=a_{6}=0$. Thus, on the basis of Eq.~(\ref{eq_nrm_1}), we obtain equation, which relates the quantities $\sigma_{y}$, $T_{g}$ and $E$:
\begin{equation}\label{eq_nrm_2}
E(\sigma_{y},T_{g})=25+41.4\sigma_{y}-0.0046T_{g}+0.0015\sigma_{y}T_{g}.
\end{equation}
In geometry, equation of this kind is known as the equation of a planar surface with a certain tilt.
\begin{figure}[h!]
	\centering
	\includegraphics[width=1.0\linewidth]{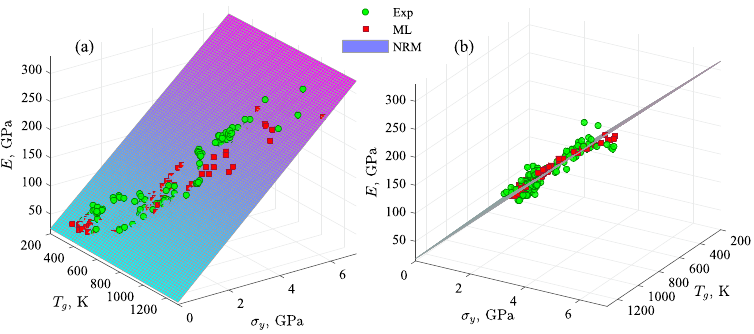}
	\caption{(a) Correspondence between the Young's modulus $E$, the yield stress $\sigma_{y}$ and the glass transition temperature $T_{g}$. The experimental values of $E$ (Exp) are compared with the Young's modulus predicted based on machine learning (ML). The result of a non-linear regression model (NRM) obtained by Eq~(\ref{eq_nrm_2}) is presented as a planar surface. (b) This correspondence from a different angle showing that all of the data are located in the vicinity of the planar surface.}
	\label{fig_5}
\end{figure}

Figure~\ref{fig_5} shows that the experimental and predicted values of the Young's modulus are indeed located on the planar surface either in its immediate vicinity; this surface is reproduced by Eq.~(\ref{eq_nrm_2}). At the same time, the data predicted on the basis of machine learning are located closer to the planar surface shown in Figure~\ref{fig_5} compared to the experimental data. The experimental data are also located along this surface and follow its slope, whereas these data have a much larger scatter compared to machine learning predictions. For example, a significant divergence between the experimental data and the result of Eq.~(\ref{eq_nrm_2}) is observed in the case of alloys based on tungsten and cobalt containing up to 32\% of boron. Here, it should be noted that the glass transition temperature $T_{g}$ included in Eq.~(\ref{eq_nrm_2}) is not fixed for a specific alloy and, as a rule, depends on the cooling rate of the liquid melt. The yield stress $\sigma_{y}$ also depends on the mechanical test conditions such as the strain rate as well as the direction and type of the applied force. Therefore, the absence of unified preparation and deformation protocols for various types of amorphous alloys may be one of the reasons for the divergence between the experimental data and the result of regression analysis. 

\section{Conclusions}

Thus, based on a large set of empirical data for amorphous metal alloys of various compositions, a relationship has been established between the Young's modulus $E$ and such key physical characteristics of alloys as the molar mass $M$, the number of components $n$, the yield stress $\sigma_{y}$, the glass transition temperature $T_{g}$. Using the machine learning based on the ANN's, it was found that the values of the Young's modulus of metal alloys under normal conditions correlate with $\sigma_{y}$ and $T_{g}$. As found, a non-linear regression model correctly reproduces the experimental data  $E$, $T_{g}$ and $\sigma_{y}$ for different metal alloys, and this model generalizes the previously known empirical relation $E/\sigma_ {y}\simeq49.8$ for amorphous metal alloys~\cite{Qu_Liu_2015}. At the same time, the results reveal that the molar mass $M$ and the number of components $n$ are not significant factors and do not affect on values of the Young's modulus.

The obtained results are of great practical importance, since they confirm the possibility of estimating the mechanical properties of metal alloys based on information about other physical properties, which can be used in the synthesis of alloys with a required properties. Moreover, the combination of the ANN and regression analysis can be applied to solve other similar tasks, where it is required to predict the physical parameters of various materials based on a limited set of training data~\cite{Bundela_Rahul_2022}. For example, such tasks include determining the GFA of various types of materials~\cite{Rickman_Kalinin_2019,Tan_Liang_2022}, as well as the task of predicting the low-temperature characteristics of glass-formers (glass transition temperature, fragility index, etc.) from high-temperature data (viscosity, melting temperature, Arrhenius transition temperature, etc.)~\cite{Balyakin_Yuryev_2022,Jaiswal_Egami_2016,Ashcraft_Kelton_2018,Xiong_Shi_2020}. 

\section*{Acknowledgement}
This work was supported by the Russian Science Foundation (project no. 19-12-00022).


\begin{thebibliography}{100}

\bibitem{Pelleg_book_2013}
J. Pelleg, Mechanical Properties of Materials, Springer Dordrecht, Netherlands, 2013.

\bibitem{Preetha_Sreekala_2018}
B. Preetha, M.S. Sreekala, T. Sabu, Fundamental Biomaterials: Metals, A volume in Woodhead Publishing Series in Biomaterials, Woodhead Publishing, Great Britain, 2018. 

\bibitem{Cavaliere_2021}
P. Cavaliere, Fatigue and Fracture of Nanostructured Materials,  Springer International Publishing, Springer, Germany, 2021.

\bibitem{Zhuang_Liu_2012}
Y.X. Zhuang, W.J. Liu, Z.Y. Chen, H.D. Xue, J.C. He, Effect of elemental interaction on microstructure and mechanical properties of FeCoNiCuAl alloys, Materials Science and Engineering: A 556 (2012) 395--399. https://doi.org/10.1016/j.msea.2012.07.003

\bibitem{Dai_Yu_Dong_2022}
H. Dai, M. Yu, Y. Dong, W. Setyawan, N. Gao, X. Wang, Effect of Cr and Al on Elastic Constants of FeCrAl Alloys Investigated by Molecular Dynamics Method, Metals 12 (2022) 558.
https://doi.org/10.3390/met12040558

\bibitem{Galimzyanov_Doronina_2022}
B.N. Galimzyanov, M.A. Doronina, A.V. Mokshin, Unusual effect of high pressures on phase transformations in Ni$_{62}$Nb$_{38}$ alloy, Journal of Physics and Chemistry of Solids 171 (2022) 110995. https://doi.org/10.1016/j.jpcs.2022.110995

\bibitem{Wang_2006}
W.H.~Wang, Correlations between elastic moduli and properties in bulk metallic glasses, J. Appl. Phys. 99 (2006) 093506. https://doi.org/10.1063/1.2193060

\bibitem{Torres_Stafford_2010}
J.M. Torres, C.M. Stafford, B.D. Vogt, Impact of molecular mass on the elastic modulus of thin polystyrene films, Polymer 51 (2010) 4211--4217. http://dx.doi.org/10.1016/j.polymer.2010.07.003

\bibitem{Binder_Kob_2005}
K. Binder, W. Kob, Glassy Materials and Disordered Solids: An Introduction to Their
Statistical Mechanics, World Scientific, Singapore, 2005.

\bibitem{Wang_Angell_2006}
L.-M. Wang, C.A. Angell, R. Richert, Fragility and thermodynamics in
nonpolymeric glass-forming liquids, J. Chem. Phys. 125 (2006) 074505, https://
doi.org/10.1063/1.2244551.

\bibitem{Chahal_Ramesh_2022}
S. Chahal, K.Ramesh, Glass formation, thermal stability and fragility minimum in Ge-Te-Se glasses, Materials Research Bulletin 152 (2022) 111833. https://doi.org/10.1016/j.materresbull.2022.111833

\bibitem{LouzguineLuzgin_2022}
D.V. Louzguine-Luzgin, Structural Changes in Metallic Glass-Forming Liquids on Cooling and Subsequent Vitrification in Relationship with Their Properties, Materials 15 (2022) 7285. https://doi.org/10.3390/ma15207285 

\bibitem{Baggioli_Zaccone_2022}
M. Baggioli, M. Landry, A. Zaccone, Deformations, relaxation, and broken symmetries in liquids, solids, and glasses: A unified topological field theory, Phys. Rev. E 105 (2022) 024602. https://doi.org/10.1103/PhysRevE.105.024602

\bibitem{Li_2001}
Y. Li, A relationship between glass-forming ability and reduced glass transition temperature near eutectic composition, Materials Transactions 42 (2001) 556--561. https://doi.org/10.2320/matertrans.42.556

\bibitem{Masood_Belova_2020}
A. Masood, L. Belova, V. Str\"{o}m, On the correlation between glass forming ability (GFA) and soft magnetism of Ni-substituted Fe-based metallic glassy alloys, Journal of Magnetism and Magnetic Materials 504 (2020) 166667. https://doi.org/10.1016/j.jmmm.2020.166667

\bibitem{Galimzyanov_Mokshin_viscosity_2021}
B.N. Galimzyanov, A.V. Mokshin, A novel view on classification of glass-forming liquids and empirical viscosity model, Journal of Non-Crystalline Solids 570 (2021) 121009.  https://doi.org/10.1016/j.jnoncrysol.2021.121009	

\bibitem{Beghini_Bertini_2006}
M. Beghini, L. Bertini, V. Fontanari, Evaluation of the stress–strain curve of metallic materials by spherical indentation, International Journal of Solids and Structures 43 (2006) 2441--2459. https://doi.org/10.1016/j.ijsolstr.2005.06.068

\bibitem{Galimzyanov_Mokshin_NiTi_2021}
B.N. Galimzyanov, A.V. Mokshin, Mechanical response of mesoporous amorphous NiTi alloy to external deformations, International Journal of Solids and Structures 224 (2021) 111047.  https://doi.org/10.1016/j.ijsolstr.2021.111047

\bibitem{Clickner_Ekin_2006}
C.C. Clickner, J.W. Ekin, N. Cheggour, C.L.H. Thieme, Y. Qiao, Y.-Y. Xie, A. Goyal, Mechanical properties of pure Ni and Ni-alloy substrate materials for Y–Ba–Cu–O coated superconductors, Cryogenics 46 (2006) 432--438. https://doi.org/10.1016/j.cryogenics.2006.01.014

\bibitem{Arrayago_Gardner_2015}
I. Arrayago, E. Real, L. Gardner, Description of stress-strain curves for stainless steel alloys, Mater. Des. 87 (2015) 540--552. https://doi.org/10.1016/j.matdes.2015.08.001

\bibitem{Liu_Zhao_2022}
X. Liu, P. Xu, J. Zhao, W. Lu, M. Li, G. Wang, Material machine learning for alloys: Applications, challenges and perspectives, Journal of Alloys and Compounds 921 (2022) 165984. https://doi.org/10.1016/j.jallcom.2022.165984 

\bibitem{Mokshin_Mirziyarova_2020}
A.V. Mokshin, V.V. Mokshin, D.A. Mirziyarova, Formation of Regression Model for Analysis of Complex Systems Using Methodology of Genetic Algorithms, Nonlinear Phenomena in Complex Systems 23 (2020) 317-326. https://doi.org/10.33581/1561-4085-2020-23-3-317-326

\bibitem{Klimenko_Ryltsev_2022}
D. Klimenko, N. Stepanov, R. Ryltsev, S. Zherebtsov, Phase prediction in high-entropy alloys with multi-label artificial neural network, Intermetallics 151 (2022) 107722. https://doi.org/10.1016/j.intermet.2022.107722

\bibitem{Mokshin_Khabibullin_2022}
A.V. Mokshin, R.A. Khabibullin, Is there a one-to-one correspondence between interparticle interactions and physical properties of liquid? Physica A: Statistical Mechanics and its Applications (2022) 128297. https://doi.org/10.1016/j.physa.2022.128297

\bibitem{Balyakin_Yuryev_2022}
I.A. Balyakin, A.A. Yuryev, V.V. Filippov, B.R. Gelchinski, Viscosity of liquid gallium: Neural network potential molecular dynamics and experimental study, Computational Materials Science 215 (2022) 111802. https://doi.org/10.1016/j.commatsci.2022.111802

\bibitem{Khakurel_Taufque_2021}
H. Khakurel, M.F.N. Taufque, A. Roy, G. Balasubramanian, G. Ouyang, J. Cui, D.D. Johnson, R. Devanathan, Machine learning assisted prediction of the Young's modulus of compositionally complex alloys, Scientific Reports 11 (2021) 17149. https://doi.org/10.1038/s41598-021-96507-0

\bibitem{Yang_Xu_2019}
K. Yang, X. Xu, B. Yang, B. Cook, H. Ramos, N.M.A. Krishnan, M.M. Smedskjaer, C. Hoover, M. Bauchy, Predicting the Young's Modulus of Silicate Glasses using High Throughput Molecular Dynamics Simulations and Machine Learning, Scientific Reports 9 (2019) 8739. https://doi.org/10.1038/s41598-019-45344-3

\bibitem{Pugar_Gang_2022}
J.A. Pugar, C. Gang, C. Huang, K.W. Haider, N.R. Washburn, Predicting Young's Modulus of Linear Polyurethane and Polyurethane--Polyurea Elastomers: Bridging Length Scales with Physicochemical Modeling and Machine Learning, ACS Appl. Mater. Interfaces 14 (2022) 16568--16581. https://doi.org/10.1021/acsami.1c24715

\bibitem{Shahani_Zheng_2022}
N.M. Shahani, X. Zheng, X. Guo, X. Wei, Machine Learning-Based Intelligent Prediction of Elastic Modulus of Rocks at Thar Coalfield, Sustainability 14 (2022) 3689. https://doi.org/10.3390/su14063689

\bibitem{Qu_Liu_2015}
R.T. Qu, Z.Q. Liu, R.F. Wang, Z.F. Zhang, Yield strength and yield strain of metallic glasses and their correlations with glass transition temperature, Journal of Alloys and Compounds 637 (2015) 44--54. https://doi.org/10.1016/j.jallcom.2015.03.005

\bibitem{Chumachenko_Gabbouj_2022}
K. Chumachenko, A. Iosifidis, M. Gabbouj, Feedforward neural networks initialization based on discriminant learning, Neural Networks 146 (2022) 220--229. https://doi.org/10.1016/j.neunet.2021.11.020

\bibitem{Haykin_2009}
S. Haykin, Neural networks and learning machines, third ed., Pearson Education Inc., New Jersey, 2009.

\bibitem{Seber_Wild_1989}
G.A.F. Seber, C.J. Wild, Nonlinear Regression. New York: John Wiley and Sons, 1989.

\bibitem{Bundela_Rahul_2022}
A.S. Bundela, M.R. Rahul, Machine learning-enabled framework for the prediction of mechanical properties in new high entropy alloys, Journal of Alloys and Compounds 908 (2022) 164578. https://doi.org/10.1016/j.jallcom.2022.164578  

\bibitem{Rickman_Kalinin_2019}
J.M. Rickman, T. Lookman, S.V. Kalinin, Materials informatics: From the atomic-level to the continuum, Acta Materialia 168 (2019) 473--510. https://doi.org/10.1016/j.actamat.2019.01.051

\bibitem{Tan_Liang_2022}
B. Tan, Y.-C. Liang, Q. Chen, L. Zhang, J.-J. Ma, Discovery of a new criterion for predicting glass-forming ability based on symbolic regression and artificial neural network, Journal of Applied Physics 132 (2022) 125104. https://doi.org/10.1063/5.0105445

\bibitem{Jaiswal_Egami_2016}
A. Jaiswal, T. Egami, K.F. Kelton, K.S. Schweizer, and Y. Zhang, Correlation between Fragility and the Arrhenius Crossover Phenomenon in Metallic, Molecular, and Network Liquids, Phys. Rev. Lett. 117 (2016) 205701. https://doi.org/10.1103/PhysRevLett.117.205701

\bibitem{Ashcraft_Kelton_2018}
R. Dai, R. Ashcraft, and K.F. Kelton, A possible structural signature of the onset of cooperativity in metallic liquids, Journal of Chemical Physics 148 (2018) 204502. https://doi.org/10.1063/1.5026801

\bibitem{Xiong_Shi_2020}
J. Xiong, S.-Q. Shi, T.-Y. Zhang, A machine-learning approach to predicting and understanding the properties of amorphous metallic alloys, Materials \& Design 187 (2020) 108378. https://doi.org/10.1016/j.matdes.2019.108378

\end{thebibliography}
\end{document}